\documentclass[letter]{aa}

\usepackage{graphicx}
\usepackage{txfonts}
\usepackage{amsmath}
\usepackage[hyperfootnotes=false,hidelinks]{hyperref}

\graphicspath{{figures/}}

\providecommand{\nolinenumbers}{}

\begin{document}

\title{Disk dispersal freezes overstable resonant librations}

\author{Linghong Lin\inst{1} \and Beibei Liu\inst{1 \star}}

\institute{Institute for Astronomy, School of Physics, Zhejiang University, Hangzhou 310027, China\\
              \email{bbliu@zju.edu.cn}}

\date{}

\abstract
{Convergent migration in a gaseous protoplanetary disk can capture a planet pair into mean-motion resonance. Eccentricity damping can subsequently make the resonant libration overstable and drive the pair out of resonance. Most studies of this process, however, assumed a static disk.}
{We examine how the decay of disk torques during dispersal changes this outcome and whether it can freeze an overstable libration before the pair escapes.}
{We describe disk dispersal by allowing the migration and eccentricity-damping timescales to increase exponentially on a local timescale $\tau_{\rm d}$. Integration of the time-dependent growth rate predicts $\tau_{\rm d,crit}\propto\tau_{e,0}$. We tested this scaling with direct $N$-body integrations and related $\tau_{\rm d}$ to the time taken by a photoevaporative cavity edge to cross the local torque-producing region.}
{The simulations recovered a linear boundary, $\tau_{\rm d,crit}\simeq S\tau_{e,0}$, with $S\simeq3$ over the explored parameter range. In a fiducial minimum-mass solar nebula, faster propagation of the cavity edge shortened the local dispersal time. The ratio $\tau_{\rm d}/\tau_{\rm d,crit}$ also decreased with orbital radius, which means that both effects favour resonant survival.}
{When local disk dispersal is sufficiently rapid, the libration amplitude can freeze and the planet pair can remain in resonance instead of escaping through overstability. Late disk evolution can therefore alter the outcome of resonant overstability.}


\maketitle
\nolinenumbers

\section{Introduction}
\label{sec:introduction}

Low-mass planets embedded in protoplanetary disks exchange angular momentum
with the surrounding gas and undergo Type I migration
\citep{KleyNelson2012,Paardekooper2023}. When two planets migrate
convergently, their period ratio decreases, and they can be captured into a
mean-motion resonance (MMR; \citealt{LeePeale2002,PapaloizouSzuszkiewicz2005}).
Resonant capture is therefore a natural outcome of disk-driven migration.

Capture does not, however, guarantee permanent resonant retention.
Eccentricity damping changes the equilibrium that is reached after capture and can
also control its stability. Under some conditions, the dissipative resonant
equilibrium becomes overstable: small librations grow until the trajectory
crosses the separatrix and the planets escape from resonance
\citep{goldreich2014,Deck2015}. Building on these studies,
\citet[hereafter Paper~I]{Lin2025} developed a unified framework for resonant
capture, stability, and overstable escape under laminar Type I migration.
This framework was subsequently extended to turbulent disks, where stochastic
forcing can further destabilise resonant systems
\citep{Nelson2005,BatyginAdams2017,Wu2024,Lin2026Turbulent}.

Previous studies of resonant overstability have generally focused on
non-dispersing disks. In these models, the migration and damping forces can
depend on the orbital elements or be obtained from hydrodynamic simulations,
while secular disk depletion is not included explicitly. During the final
clearing phase, however, photoevaporation can open an inner cavity and drive rapid
inside-out dispersal \citep{Owen2012,Ercolano2021}. As the local gas density
falls, the migration and eccentricity-damping timescales lengthen, and overstable
growth slows. Consequently, the local disk forcing might fade before the libration
reaches the separatrix, leaving the planets in resonance.

Motivated by this possibility, we examine in this Letter the finite-growth
mechanism by which disk dispersal limits the accumulated overstable growth. We
first derive the linear scaling of the critical local disk-dispersal time with
the initial eccentricity-damping time and test it with direct $N$-body
integrations in Sect.~\ref{sec:model}. We then relate the local dispersal time to
photoevaporative inside-out clearing and determine where resonant survival is
possible in a fiducial disk in Sect.~\ref{sec:disk-clearing}. Finally,
Sect.~\ref{sec:summary} summarises our conclusions.

\section{Finite growth during disk dispersal}
\label{sec:model}
We first derive the linear scaling of the critical local disk-dispersal
timescale $\tau_{\rm d,crit}$ with the initial eccentricity-damping timescale
$\tau_{e,0}$ in Sect.~\ref{sec:theory}. We then test this scaling with direct
\(N\)-body simulations in Sect.~\ref{sec:nbody}.

\subsection{Finite-growth scaling}
\label{sec:theory}

\begin{figure*}
 \centering
 \includegraphics[width=0.49\textwidth]{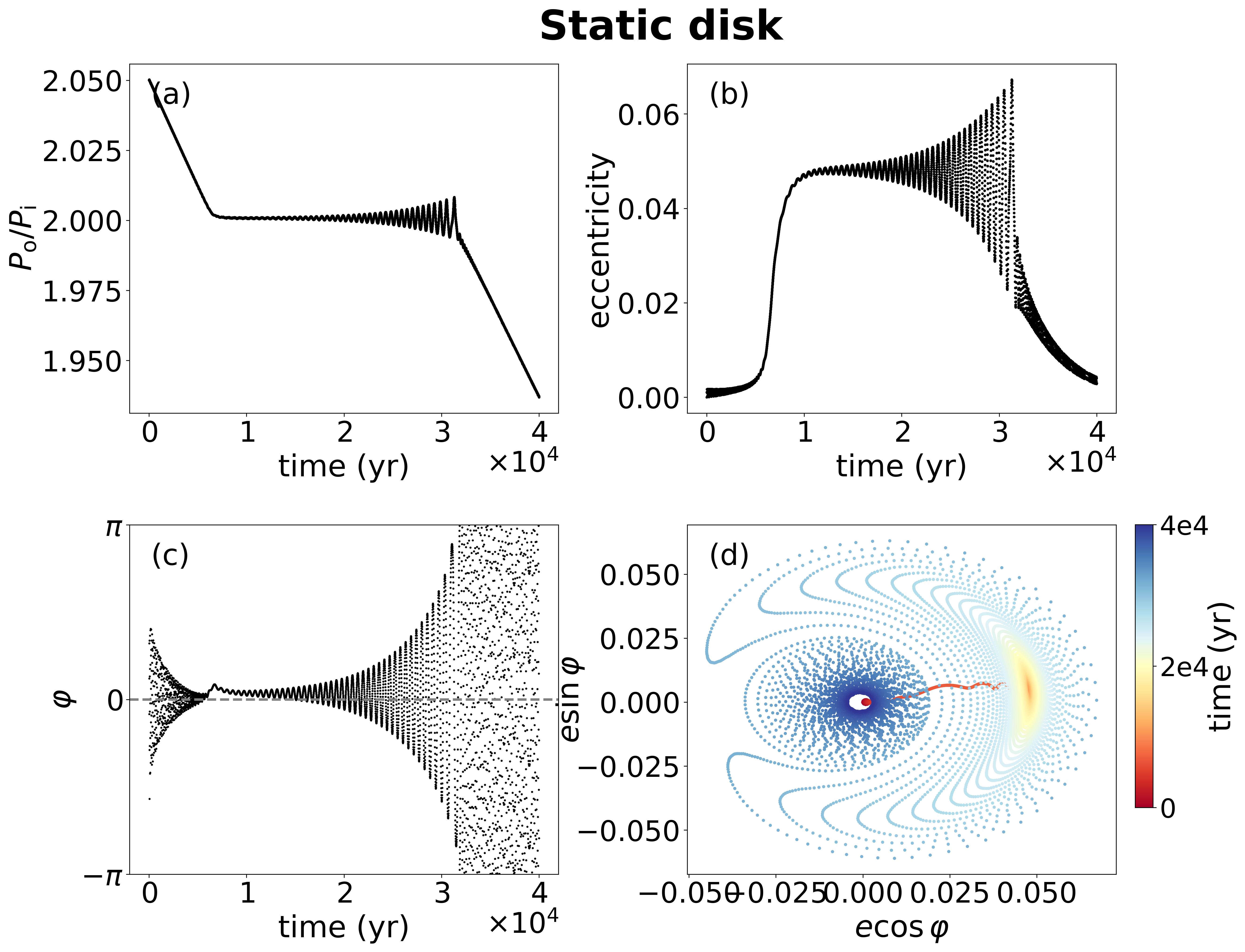}\hfill
 \includegraphics[width=0.49\textwidth]{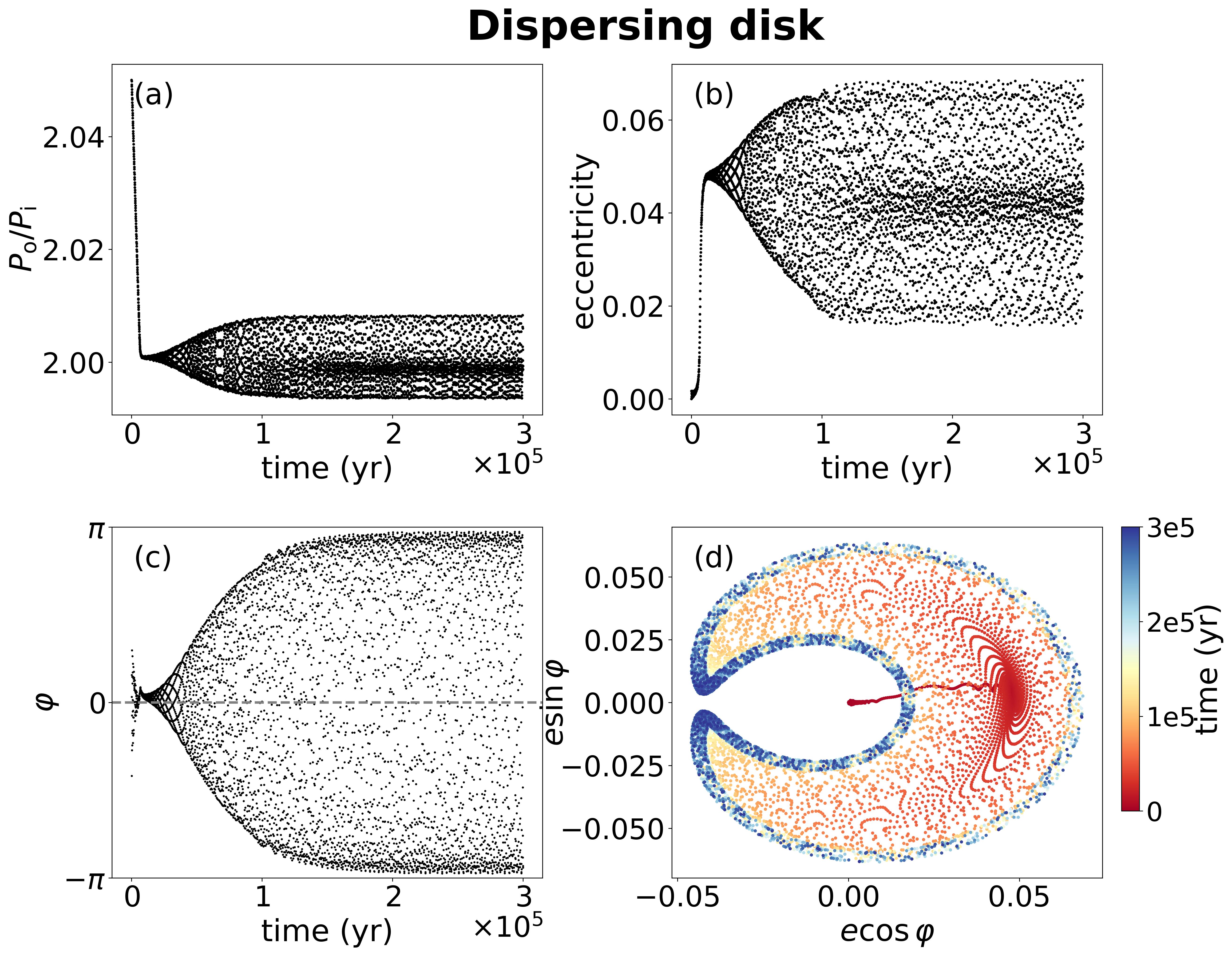}
 \caption{Resonant evolution in static and dispersing disks. The left
 and right sets of panels show the same planet pair in a static and dispersing
 disk, respectively. Within each set, the panels show the period ratio,
 inner-planet eccentricity, resonant angle, and phase-space trajectory. Both
 cases adopt $\tau_{\rm m,0}=8\times10^5\,\mathrm{yr}$,
 $\tau_{\rm m,0}/\tau_{e,0}=200$, and
 $\tau_{e,0}=4\times10^3\,\mathrm{yr}$. The static case escapes after crossing
 the separatrix, whereas dispersal with
 $\tau_{\rm d}=2.3\times10^4\,\mathrm{yr}$ freezes the libration at a finite
 amplitude and preserves the resonance.}
 \label{fig:timeseries}
\end{figure*}

For the analytical model, we considered the limit in which an inner low-mass
planet is captured into a first-order \(j:j-1\) MMR with a much more massive
outer planet. This corresponds to the dissipative circular restricted
three-body approximation. The relevant resonant angle is
\begin{equation}
    \varphi=j\lambda_{\rm o}-(j-1)\lambda_{\rm i}-\varpi_{\rm i},
\end{equation}
where \(\lambda_{\rm i}\) and \(\lambda_{\rm o}\) are the mean longitudes
of the inner and outer planets, and \(\varpi_{\rm i}\) is the longitude
of pericentre of the inner planet. The subscripts i and o refer
to the inner and outer planets throughout.

At low eccentricity, planet--disk dissipation can be parametrised as
\citep{Teyssandier2014,Ataiee2021}
\begin{subequations} \label{disk_ef}
\begin{eqnarray}
      \frac{1}{L}\frac{dL}{dt} & = & -\frac{1}{\tau_{\rm m}}, \\
      \frac{1}{e}\frac{de}{dt} & = & -\frac{1}{\tau_e},
      \label{de_dt}
\end{eqnarray}
\end{subequations}
where \(L\) and \(e\) are the orbital angular momentum and eccentricity
of the planet. The timescales \(\tau_{\rm m}\) and \(\tau_e\) characterise
angular-momentum and eccentricity damping, respectively.

Following resonant capture, eccentricity damping can destabilise the
dissipative resonant equilibrium, causing the libration amplitude to
grow through resonant overstability
\citep{goldreich2014,Deck2015,Lin2025}.
Let \(\mathcal{A}(t)\) denote the slowly varying libration amplitude of
\(\varphi\). During the linear overstability stage,
\begin{equation}
 \frac{\dot{\mathcal A}}{\mathcal A}=\Gamma(t)
 =\frac{\gamma_j}{\tau_e(t)},
 \label{eq:growth-rate}
\end{equation}
where \(\Gamma\) is the growth rate, and \(\gamma_j\) is dimensionless
\citep{goldreich2014,Batygin2026B}.

During disk dispersal, the decreasing gas density weakens both migration
and eccentricity damping. We modelled this evolution as
\begin{equation}
 \tau_{\rm m}(t)=\tau_{\rm m,0}\exp\left(\frac{t}{\tau_{\rm d}}\right),
 \qquad
 \tau_e(t)=\tau_{e,0}\exp\left(\frac{t}{\tau_{\rm d}}\right),
 \label{eq:damping-evolution}
\end{equation}
where \(t=0\) marks the onset of dispersal, \(\tau_{\rm m,0}\) and
\(\tau_{e,0}\) are their values at that time, and \(\tau_{\rm d}\) characterises the exponential decline
of the planet-disk interaction. We assumed that migration and eccentricity
damping weaken on the same timescale, so that
\(\tau_{\rm m}/\tau_e\) remains constant. We adopt this prescription
for the numerical calculations below. Its physical interpretation in terms of
local disk clearing is discussed in Sect.~\ref{sec:disk-clearing}.

Equations~(\ref{eq:growth-rate}) and
(\ref{eq:damping-evolution}) give
\(\Gamma(t)=\Gamma_0\exp(-t/\tau_{\rm d})\), where
\(\Gamma_0\equiv\gamma_j/\tau_{e,0}\). The decay of \(\Gamma\) limits
the total growth of the libration amplitude. More generally, the
accumulated growth can be expressed in terms of an effective local dispersal
time,
\[
 \tau_{\rm eff}\equiv\int_0^\infty
 \frac{\tau_{e,0}}{\tau_e(t)}\,dt .
\]
The final libration amplitude is then
\begin{equation}
 \mathcal A_\infty
 =
 \mathcal A_0
 \exp\left[\int_0^\infty \Gamma(t)\,dt\right]
 =
 \mathcal A_0\exp\left(\frac{\gamma_j\tau_{\rm eff}}
 {\tau_{e,0}}\right),
 \label{eq:final-amplitude}
\end{equation}
where \(\mathcal A_0\) is the amplitude at the onset of dispersal, and
\(\mathcal A_\infty\) is the final frozen amplitude. For the
exponential prescription in Eq.~(\ref{eq:damping-evolution}),
\(\tau_{\rm eff}=\tau_{\rm d}\), so \(\tau_{\rm d}\) is used as the dispersal
parameter in the simulations below. Disk dispersal
therefore limits the total overstable growth whenever \(\tau_{\rm eff}\) is
finite.

Escape occurs when the libration amplitude reaches a critical value and
the resonant angle changes from libration to circulation. We denote this
threshold by \(\mathcal A_{\rm esc}\), with
\(\mathcal A_{\rm esc}\sim\pi\) as an order-of-magnitude estimate.
The condition \(\mathcal A_\infty<\mathcal A_{\rm esc}\) therefore
gives the general survival criterion
\(\tau_{\rm eff}<S\tau_{e,0}\), where
\(S\equiv\gamma_j^{-1}\ln(\mathcal A_{\rm esc}/\mathcal A_0)\).
For the exponential prescription adopted here, this reduces to a
critical dispersal time
\begin{align}
 \tau_{\rm d,crit}
 =
 \frac{1}{\gamma_j}
 \ln\left(\frac{\mathcal A_{\rm esc}}{\mathcal A_0}\right)
 \tau_{e,0}.
 \label{eq:critical-time}
\end{align}
Thus, \(\tau_{\rm d,crit}=S\tau_{e,0}\), where \(S\) is a dimensionless
coefficient that depends on the planet--resonance
configuration and on the migration and damping history. We do not
attempt to derive its full dependence here. Instead, we use the analytic
model to predict the linear scaling with \(\tau_{e,0}\) and determine
\(S\) numerically for the system considered below.

The corresponding survival condition is
\begin{equation}
 \tau_{\rm d}<\tau_{\rm d,crit}=S\tau_{e,0}.
 \label{eq:survival-condition}
\end{equation}
Figure~\ref{fig:timeseries} shows the contrast: in a static disk, the
libration grows through the separatrix and the pair escapes; in a dispersing
disk, the weakening forcing freezes the libration below the separatrix and the
pair remains resonant.
The argument applies to other overstable MMRs provided that their slow
amplitude evolution follows the same form, although \(S\) must be recalibrated.

\subsection{Numerical test of the scaling}
\label{sec:nbody}

We tested Eq.~(\ref{eq:critical-time}) with \texttt{REBOUND}/\texttt{REBOUNDx}
integrations \citep{ReinLiu2012,Tamayo2020}. A solar-mass star hosts two
coplanar planets with \(m_{\rm i}=1\,M_\oplus\) and
\(m_{\rm o}=10\,M_\oplus\) near the 2:1 MMR. Only the outer planet migrates
inward, while the two eccentricities are damped on the same \(\tau_e\). For each
\(\tau_{\rm m,0}=[4,8,12]\times10^5\,\mathrm{yr}\), a static reference run
set the dispersal onset when the resonant angle had converged. We then varied
\(\tau_{e,0}\) and scanned \(\tau_{\rm d}\) for the boundary between separatrix
crossing and resonant survival. Appendix~\ref{app:numerics} gives the full
numerical setup.

\begin{figure}
 \centering
 \includegraphics[width=\linewidth]{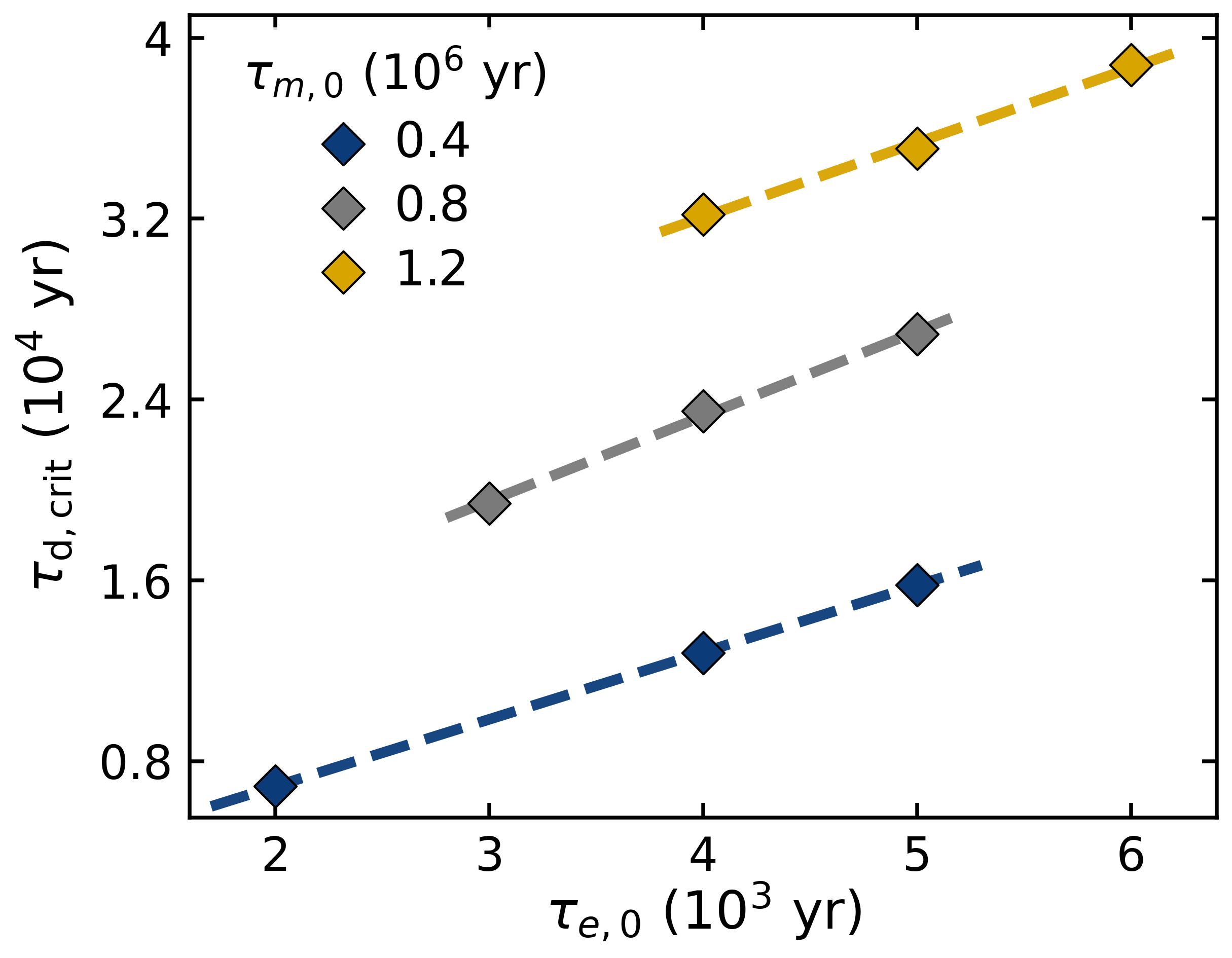}
 \caption{Critical dispersal time vs. initial
 eccentricity-damping time. The systems are near the 2:1 resonance, with
 $m_{\rm i}=1\,M_\oplus$ and $m_{\rm o}=10\,M_\oplus$. The symbols mark the
 numerical survival--escape boundary for three values of $\tau_{\rm m,0}$.
 The dashed lines show linear fits with slopes of the order of three.}
 \label{fig:critical-boundary}
\end{figure}

Figure~\ref{fig:critical-boundary} shows that \(\tau_{\rm d,crit}\) increases
approximately linearly with \(\tau_{e,0}\) for all three values of
\(\tau_{\rm m,0}\), confirming the predicted scaling. Fits of
\(\tau_{\rm d,crit}=S\tau_{e,0}+b\) give slopes of the order of three, with no clear
monotonic dependence on \(\tau_{\rm m,0}\); the coefficients are listed in
Table~\ref{tab:critical-fits}. We adopt the representative value \(S=3\) below.

\section{Connection to local disk clearing}
\label{sec:disk-clearing}
In Sect.~\ref{sec:physical-scenario} we relate \(\tau_{\rm d}\) to the local
decay of disk torques during photoevaporative inside-out clearing. In
Sect.~\ref{sec:parameter-dependence} we compare it with
\(\tau_{\rm d,crit}\) in an illustrative Type I disk. The fits in
Sect.~\ref{sec:nbody} motivate \(S=3\) as a representative value over the
explored parameter range, giving \(\tau_{\rm d,crit}=3\tau_{e,0}\).
Figure~\ref{fig:physical-feasibility} therefore provides an illustrative map
of the survival regime and not a precise boundary.

\subsection{Local clearing timescale}
\label{sec:physical-scenario}
High-energy stellar radiation drives photoevaporative winds. When the disk
accretion rate falls below the wind mass-loss rate, the inner disk drains and
a cavity opens; continued mass loss drives its edge outward
\citep{Owen2010,Owen2012,ErcolanoPascucci2017,Picogna2019,Liu2022,Ying2026}.
As this edge crosses the torque-producing region around a planet, the local
surface density falls, and migration and eccentricity damping both weaken.
For the sharp-edge picture considered here, we interpreted the
exponential dispersal time \(\tau_{\rm d}\) as the time required for the cavity
edge to cross the local torque-producing region. Most of the
Lindblad torque on a low-mass planet is generated within a few pressure scale
heights
\citep{Dangelo2010,Cimerman2024,Brown2024,Wu2024}. We therefore adopted
\(\Delta\simeq H\), where \(H\) is the local pressure scale height, giving
\begin{equation}
 \tau_{\rm d}\simeq\frac{H}{v_{\rm edge}},
 \label{eq:local-clearing}
\end{equation}
where \(v_{\rm edge}=dR_{\rm edge}/dt\) is the local edge speed. We explored
\(v_{\rm edge}=10\)--\(10^3\,\mathrm{au\,Myr^{-1}}\), which is consistent with
 photoevaporative estimates \citep{Liu2022}; Appendix~\ref{app:edge-speed}
 gives the normalisation. Equation~(\ref{eq:local-clearing}) is
 therefore an order-of-magnitude sharp-edge mapping of the exponential decay
 law adopted in Sect.~\ref{sec:model} and not a general definition of the
 dispersal time.

\subsection{Survival during local disk clearing}
\label{sec:parameter-dependence}

The relevant question is whether the local disk torque fades before overstable
librations carry the planets across the separatrix. We examined this condition
in an MMSN disk around a solar-mass star \citep{Hayashi1981},
\begin{equation}
\Sigma_g=1700\left(\frac{a}{1\,\mathrm{au}}\right)^{-3/2}
\mathrm{g\,cm^{-2}},\quad
h\equiv\frac{H}{a}=0.033
\left(\frac{a}{1\,\mathrm{au}}\right)^{1/4}.
\label{eq:mmsn}
\end{equation}
For a low-mass planet of mass \(M_p\) on a nearly circular orbit, linear Type I
theory gives \citep{TanakaWard2004}
\begin{align}
\tau_e={}&\frac{1}{0.780}
\left(\frac{M_\star}{M_p}\right)
\left(\frac{M_\star}{\Sigma_g a^2}\right)
h^4\Omega^{-1} \nonumber\\
\simeq{}&4.2\times10^2\,\mathrm{yr}
\left(\frac{M_\oplus}{M_p}\right)
\left(\frac{a}{1\,\mathrm{au}}\right)^2.
\label{eq:typeI-taue}
\end{align}
Here, \(M_\star\) is the stellar mass, and \(\Omega\) is the Keplerian
frequency. We identified this damping time at the onset of dispersal with
\(\tau_{e,0}\). The critical time therefore scales as
\(\tau_{\rm d,crit}=3\tau_{e,0}\propto M_p^{-1}a^2\).

The competing clearing time follows from the passage of the cavity edge.
Using \(\Delta\simeq H\) in Eq.~(\ref{eq:local-clearing}) gives
\begin{equation}
\tau_{\rm d}\simeq3.3\times10^2\,\mathrm{yr}
\left(\frac{100\,\mathrm{au\,Myr^{-1}}}{v_{\rm edge}}\right)
\left(\frac{a}{1\,\mathrm{au}}\right)^{5/4}.
\label{eq:taud-scaling}
\end{equation}
Their ratio is
\begin{align}
\mathcal R\equiv\frac{\tau_{\rm d}}{\tau_{\rm d,crit}}
\simeq{}&2.6
\left(\frac{M_p}{10\,M_\oplus}\right)
\left(\frac{100\,\mathrm{au\,Myr^{-1}}}{v_{\rm edge}}\right)
\left(\frac{a}{1\,\mathrm{au}}\right)^{-3/4}.
\label{eq:timescale-ratio}
\end{align}
Resonant survival corresponds to \(\mathcal R<1\).

\begin{figure}
 \centering
 \includegraphics[width=1\columnwidth]{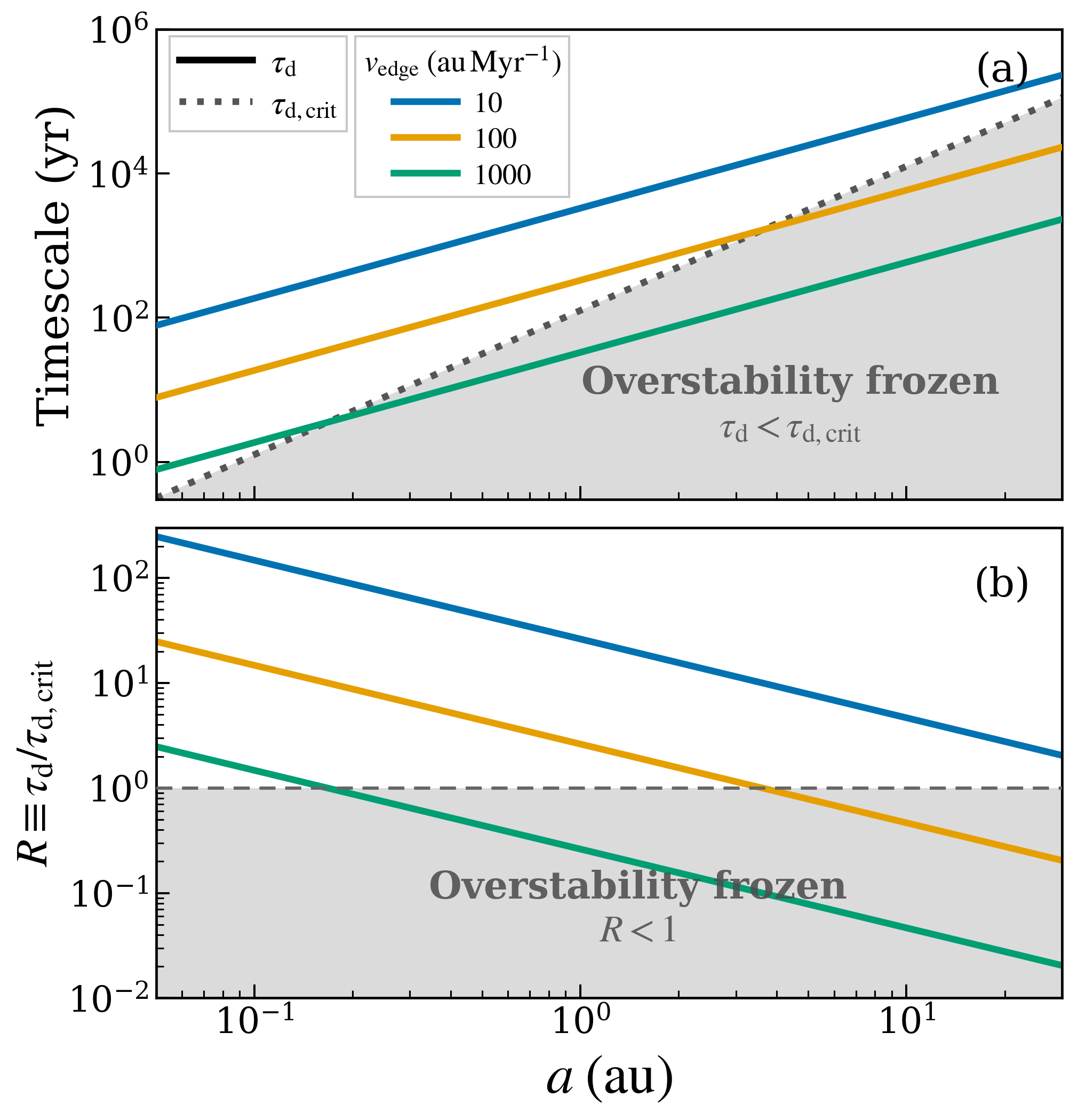}
 \caption{Illustrative sharp-edge comparison of critical and local clearing
 times. The planet mass is $M_p=10\,M_\oplus$. The top panel compares
 $\tau_{\rm d,crit}=3\tau_{e,0}$ with $\tau_{\rm d}=H/v_{\rm edge}$. The
 bottom panel shows $\mathcal R=\tau_{\rm d}/\tau_{\rm d,crit}$. The grey regions
 mark $\mathcal R<1$, where libration growth can be frozen before escape.}
 \label{fig:physical-feasibility}
\end{figure}

Figure~\ref{fig:physical-feasibility} illustrates this comparison for a
representative planet with \(M_p=10\,M_\oplus\), with the grey regions marking
\(\mathcal R<1\). Because \(\tau_{\rm d,crit}\) increases with orbital radius
more rapidly than \(\tau_{\rm d}\), \(\mathcal R\) decreases outward;
faster-moving edges also reduce \(\tau_{\rm d}\). Thus, resonant survival is
favoured at larger orbital radii and for faster-moving edges. For the explored
edge speeds, the survival boundary lies within the Type I regime shown,
indicating that the mechanism can operate under the fiducial assumptions.
Equation~(\ref{eq:timescale-ratio}) also suggests that lower-mass planets are
easier to retain when \(S\) is fixed.

The survival boundary also depends on when dispersal begins relative to
resonance capture. In our simulations, dispersal begins shortly after capture,
when the resonant configuration has been established and while the libration
amplitude remains small. If
dispersal begins only after overstability has amplified the libration, less
additional growth is needed for escape and the corresponding
\(\tau_{\rm d,crit}\) is smaller. Within the fiducial disk model, survival
would then require faster clearing or larger orbital radii. By
contrast, a pair that forms or reaches resonance only after the disk has
already become depleted experiences weaker eccentricity damping and hence a
larger \(\tau_{\rm d,crit}\), which favours survival.

The boundary in Fig.~\ref{fig:physical-feasibility} assumes a
well-defined outward-moving cavity edge. In reality, photoevaporative clearing
may be more complex: radial transport can partly refill a density depression
and produce a broad, partially depleted region \citep{Weber2026}, while sharp
density gradients can directly modify migration and resonant dynamics. Such
gradients might suppress overstability \citep{Batygin2026B}, whereas wave
reflection, magnetic structure, or rebound migration might slow, halt, or reverse
migration \citep{Liu2017Rebound,Liu2025Rebound,Liu2026Rebound,MirandaLai2018,CevallosSotoZhu2026}.
In such cases, \(H/v_{\rm edge}\) no longer provides a direct estimate of the
local dispersal history. For more general damping histories, the accumulated
overstable growth is instead characterised by \(\tau_{\rm eff}\) defined in
Sect.~\ref{sec:model}. Additional edge-induced changes to migration or resonant
dynamics can further shift the quantitative boundary in
Fig.~\ref{fig:physical-feasibility} without altering the basic finite-growth
picture.

The finite-growth mechanism can therefore leave a planet pair in
resonance with a finite libration amplitude, similar to the configuration
inferred for TOI-216 \citep{Nesvorny2022}. After disk dispersal, subsequent
dynamical perturbations could move the surviving pair slightly away from exact
commensurability while its resonant angle continues to oscillate, possibly
accounting for young near-resonant systems \citep{Hu2025,Wang2025}. Disk dispersal can therefore
affect both resonance survival and the configuration observed after the gas has
disappeared.

\section{Summary}
\label{sec:summary}

Disk dispersal causes the overstability growth rate to decline with time, limiting
the total amplification of the resonant libration. The accumulated
growth can be characterised by an effective local dispersal time
\(\tau_{\rm eff}\), giving the general survival condition
\(\tau_{\rm eff}<S\tau_{e,0}\). For the exponential decay law adopted in our
simulations, \(\tau_{\rm eff}=\tau_{\rm d}\), reducing this condition to
Eq.~(\ref{eq:survival-condition}). Our integrations confirmed the linear scaling
and give $S\simeq3$ over the explored range
(Fig.~\ref{fig:critical-boundary}).

For a photoevaporative cavity edge, we estimated
$\tau_{\rm d}\simeq H/v_{\rm edge}$. In the fiducial MMSN model, plausible
clearing speeds cross the survival boundary in Fig.~\ref{fig:physical-feasibility},
with faster clearing and larger orbital radii favouring survival. The
quantitative boundary depends on the disk model and on $S$, but the finite-growth
argument can be applied to other overstable MMRs after recalibration. Late disk
dispersal can therefore affect whether an overstable pair escapes or remains
in resonance.

\begin{acknowledgements}
BL is supported by the National Key R\&D Program of China (2024YFA1611803),
the National Natural Science Foundation of China (Nos. 12222303 and 12173035),
and the start-up grant of the Bairen program from Zhejiang University. The
simulations and analysis presented in this article were carried out on the
SilkRiver Supercomputer of Zhejiang University.
\end{acknowledgements}

\bibliographystyle{aa}
\bibliography{main}

\clearpage
\begin{appendix}

\section{Numerical setup and fit parameters}
\label{app:numerics}

The inner planet starts at $a_{\rm i}=1\,\mathrm{au}$ with $e_{\rm i}=0$,
while the outer planet has $P_{\rm o}/P_{\rm i}=2.05$ and
$e_{\rm o}=10^{-3}$. We use the \texttt{WHFast} integrator in
\texttt{REBOUND} \citep{ReinLiu2012} and implement dissipation with
\texttt{REBOUNDx} \citep{Tamayo2020}. The accelerations are
\citep{Papaloizou2000,Cresswell2006,Cresswell2008,Pichierri2024}
\begin{equation}
 \boldsymbol{a}_{\rm m}=-\frac{\boldsymbol{v}_{\rm pl}}{\tau_{\rm m}},
 \qquad
 \boldsymbol{a}_{\rm e}=-2
 \frac{(\boldsymbol{v}_{\rm pl}\!\cdot\!\boldsymbol{r}_{\rm pl})
 \boldsymbol{r}_{\rm pl}}{r_{\rm pl}^2\tau_e}.
\end{equation}
Only the outer planet migrates inward; both eccentricities are damped on the
same $\tau_e$. We use a timestep of $0.002\,\mathrm{yr}$ and integrate for at
least $2\times10^5\,\mathrm{yr}$.

For each $\tau_{\rm m,0}$, a static-disk run first establishes the resonant
state. Dispersal begins when the resonant angle has converged towards its
equilibrium value; this occurs at $8\times10^3\,\mathrm{yr}$ in the fiducial
case in Fig.~\ref{fig:timeseries}. The onset is determined separately when
$\tau_{\rm m,0}$ changes. We classify a system as escaping if its phase-space
trajectory crosses the resonant separatrix, and as surviving otherwise.

\begin{table}[h]
 \caption{Linear fits to the survival--escape boundaries.}
 \label{tab:critical-fits}
 \centering
 \begin{tabular}{ccc}
  \hline\hline
  $\tau_{\rm m,0}$ (yr) & $S$ & $b$ (yr) \\
  \hline
  $4.00\times10^5$ & $2.964\pm0.012$ & $964\pm48$ \\
  $8.00\times10^5$ & $3.75\pm0.20$ & $8267\pm825$ \\
  $1.20\times10^6$ & $3.30\pm0.23$ & $18867\pm1170$ \\
  \hline
 \end{tabular}
 \tablefoot{The fitted relation is $\tau_{\rm d,crit}=S\tau_{e,0}+b$,
 and the boundaries are shown in Fig.~\ref{fig:critical-boundary}.
 Errors are $1\sigma$ uncertainties from the fit covariance.}
\end{table}

The fitted offset \(b\) increases over the sampled values of
\(\tau_{\rm m,0}\). At fixed \(\tau_{e,0}\), a larger
\(\tau_{\rm m,0}/\tau_{e,0}\) corresponds to stronger eccentricity damping
relative to migration and a lower resonant equilibrium eccentricity. The
system is then farther from the escape threshold and requires more growth of
the libration amplitude before leaving resonance, qualitatively allowing it to
survive for a longer \(\tau_{\rm d}\). Since only three values of
\(\tau_{\rm m,0}\) are sampled, we treat \(b\) as an empirical finite-range
offset rather than a separate scaling relation.

\section{Cavity-edge speed}
\label{app:edge-speed}

For an order-of-magnitude normalisation, we equate the photoevaporative
mass-loss rate to the gas removed as the cavity edge advances. With
$dM\simeq2\pi R_{\rm edge}\Sigma_{\rm edge}\,dR_{\rm edge}$,
\begin{align}
 v_{\rm edge}\simeq{}&
 350\,\mathrm{au\,Myr^{-1}}
 \left(\frac{\dot M_{\rm pe}}
 {10^{-9}M_\odot\,\mathrm{yr^{-1}}}\right)
 \nonumber\\
 &\times
 \left(\frac{R_{\rm edge}}{1\,\mathrm{au}}\right)^{-1}
 \left(\frac{\Sigma_{\rm edge}}
 {4\,\mathrm{g\,cm^{-2}}}\right)^{-1}.
 \label{eq:edge-normalization}
\end{align}
Here, $\dot M_{\rm pe}$ is the wind mass-loss rate and $\Sigma_{\rm edge}$ is
the surface density at the cavity edge. Both can evolve with radius and time,
so $v_{\rm edge}$ is not globally constant. We treat it as constant only while
the edge crosses the narrow local torque-producing region.

\end{appendix}

\end{document}